\documentclass{article}

\usepackage[dblblindworkshop, final]{neurips_2026}

\usepackage[utf8]{inputenc} 
\usepackage[T1]{fontenc}    
\usepackage{hyperref}       
\usepackage{url}            
\usepackage{booktabs}       
\usepackage{amsfonts}       
\usepackage{nicefrac}       
\usepackage{microtype}      
\usepackage{xcolor}         
\usepackage{graphicx}
\usepackage{amsmath}

\title{Self-Supervised Learning for Robust Resonance Mass Regression in Cascade Decays}

\author{%
  Ho Fung Tsoi\textsuperscript{1}, Alex Yang\textsuperscript{1}, Luis Felipe Gutierrez Zagazeta\textsuperscript{1}, Shion Chen\textsuperscript{2},\\\textbf{Dylan Sheldon Rankin}\textsuperscript{\textbf{1}} \\
  \textsuperscript{1}Department of Physics and Astronomy, University of Pennsylvania, USA \\
  \textsuperscript{2}Department of Physics, Kyoto University, Japan \\
  \texttt{\{hftsoi,dsrankin\}@sas.upenn.edu}
}

\begin{document}

\maketitle

\begin{abstract}
  Reconstructing the mass of a heavy resonance from its decay products with missing energy is one of the central tasks that directly determine the sensitivity in new physics searches at collider experiments. Supervised learning approaches to this problem often struggle to generalize well due to the presence of various systematic uncertainties and distribution shifts.
  Exhausting all possible variations in the labeled data can be very compute-intensive, while a failure of the model to generalize can corrupt the reconstructed resonance widths that are critical in peak-hunting analyses. In this work, following the foundation model paradigm, we use a self-supervised approach to pre-train a transformer encoder with VICReg to learn an embedding invariant to various corruptions, then fine-tune it for mass regression on a heavy resonance with masses ranging from 2.5 to 6.5 TeV and a SUSY-like cascade decay into an eleven-body final state. We show that the pre-trained model reconstructs sharper resonance peaks and has a more stable performance under various realistic corruptions, compared to a supervised model of the same architecture trained on the same augmented data from scratch.
\end{abstract}

\section{Introduction}

Searching for physics beyond the Standard Model at high-energy physics (HEP) experiments often involves reconstructing the invariant mass of hypothetical heavy resonances from their decay products, where the width of the reconstructed peak critically determines the search sensitivity.
While this is relatively straightforward for low-multiplicity final states, it becomes challenging for cascade decays of heavy resonances such as TeV-scale supersymmetric (SUSY) particles, which can result in high-multiplicity final states~\citep{ATLAS:2021fbt}.
With many final-state objects, systematic uncertainties from detector effects and object reconstruction inefficiencies can quickly compound to corrupt the reconstructed resonance width.
In these scenarios, standard supervised regression algorithms often fail to generalize beyond their training datasets, as this would require a massive amount of labeled data incorporating all kinds of systematic variations.
However, simulating such datasets in collider experiments is very compute-intensive, especially in searches with a broad hypothetical mass range and a dense mass grid, where generating sufficient variations at every single mass point is impractical for supervised learning.

Self-supervised learning (SSL) methods~\citep{9157636,chen2020simple,caron2021emerging,zhou2021ibot,bardes2022vicreg,assran2023selfsupervisedlearningimagesjointembedding,oquab2023dinov2} have been shown to be effective in HEP at generalizing across tasks and mitigating mismodeling effects compared to supervised models~\citep{Harris:2024sra,SciPostPhys.18.5.150,hao2025rino}, especially when simulated labeled data is limited.
In this work, we use SSL to first pre-train a task-agnostic embedding on data without mass labels, which learns representations invariant to systematic variations that would corrupt the reconstructed mass peak.
We then fine-tune it using mass labels to perform mass regression.
We demonstrate this approach on a dataset of heavy resonances spanning a broad TeV-scale mass range with a SUSY-like cascade decay that results in an eleven-body final state and show that the pre-trained embedding is robust against various realistic systematic variations.
When fine-tuned for mass regression, our method improves the reconstructed width across essentially all mass points and shows more stable performance against corruptions compared to a supervised model baseline.

\section{Method}

\textbf{SSL pre-training.}
Each event in our dataset is represented as a set of final-state objects, each with standard inputs such as transverse momentum ($p_{\text{T}}$), pseudorapidity ($\eta$), azimuthal angle ($\phi$), and particle identification (PID).
We use the VICReg~\citep{bardes2022vicreg} method to pre-train a transformer encoder to learn to embed events of a broad mass range into a representation space without mass labels.
Two augmented views are created per event as inputs to the pre-training architecture, and the VICReg objective is to learn invariance in the embedding between the two views as they originate from the same underlying event, with encouraging unit variance along each embedding dimension and decorrelation between pairs of embedding dimensions to avoid information collapse.
We choose the augmentations to try to emulate various realistic systematic variations so that their corruption effects on the embedding are reduced before fine-tuning it for mass regression.

\textbf{Systematic variations as augmentations.}
As listed in Tab.~\ref{tab:aug}, we independently apply seven types of augmentations per event, each with a fixed probability, to emulate different realistic systematic variations, with strengths chosen conservatively at or above the measurements~\citep{CMS:2016lmd,CMS:2017yfk,CMS:2019ctu}: (i) reconstruction inefficiency, (ii) jet energy scale, (iii) energy resolution, (iv) missing transverse energy (MET) resolution, (v) angular resolution, (vi) detector symmetry, and (vii) mis-identification.

\begin{table}[!t]
  \caption{List of augmentations for creating invariant views for VICReg pre-training, motivated by typical object performance at the LHC.}
  \label{tab:aug}
  \centering
  \small
  \scalebox{0.85}{
  \begin{tabular}{llc}
    \toprule
    Effect & Augmentation & Per-event prob. \\ \midrule
    Reco. inefficiency & Randomly drop 1-2 visible objects, whose $p_{\text{T}}$ is added to MET. & 30\% \\
    Jet energy scale & All hadronic objects have a correlated $\sigma=3\%$ shift. & 50\% \\
    Energy resolution & Each visible object has a $p_{\text{T}}$ smearing at $\sigma=10\%$. & 50\% \\
    MET resolution & MET has a $p_{\text{T}}$ smearing at $\sigma=15\%$. & 50\% \\
    Angular resolution & Each visible object has an $\eta$ smearing at $\sigma=0.05$. & 50\% \\
    Detector symmetry & All objects have a global $\phi$ rotation. & 100\% \\
    Mis-identification & A pair of hadronic objects have their PID swapped. & 15\% \\
    \bottomrule
  \end{tabular}
  }
\end{table}

\textbf{Fine-tuning for mass regression.}
For mass regression, a multilayer perceptron (MLP) head is attached to the pre-trained encoder to train on events with mass labels using a relative mean squared error loss, $\frac{1}{n}\sum_{i=1}^{n}(\frac{\hat{m}_i-m_i}{m_i})^2$, where $n$ is the number of events in the labeled dataset, $\hat{m}_i$ is the predicted mass and $m_i$ is the truth mass for the $i$-th event.
For baseline comparison, a supervised model of the same architecture as the VICReg model is used and trained on the same augmented data from scratch for mass regression.

\section{Experiments}

\textbf{Dataset.}
We use a kinematic generator~\citep{GutierrezZagazeta:2025wnc} to generate a SUSY-like cascade decay process $bA\rightarrow bgX\rightarrow bgqqqqY\rightarrow bgqqqqqqqql\nu$, where the heavy resonance $A$ decays into intermediate resonances $X$ and $Y$ and then into the final state with eleven particles: a gluon $g$, eight light quarks $q$, a charged lepton $l$, and a neutrino $\nu$.
The b-quark $b$ is excluded from the inputs for simplicity.
21 mass points for $A$ are generated in the range from 2500 GeV to 6500 GeV, in steps of 200 GeV, each with 50,000 events, totaling 1.05M events.
Each particle has inputs $(p_{\text{T}},\eta,\phi,\text{PID})$.
The neutrino is treated as MET, with $\eta=0$.
The set is split into train/val/test with ratios 0.7/0.15/0.15.
For simplicity, the data are generated at the kinematic level without detector simulation and reconstruction, and we rely on augmentations to emulate various systematic effects for training and evaluation.

\begin{figure}[!t]
    \centering
    \includegraphics[width=0.2\textwidth]{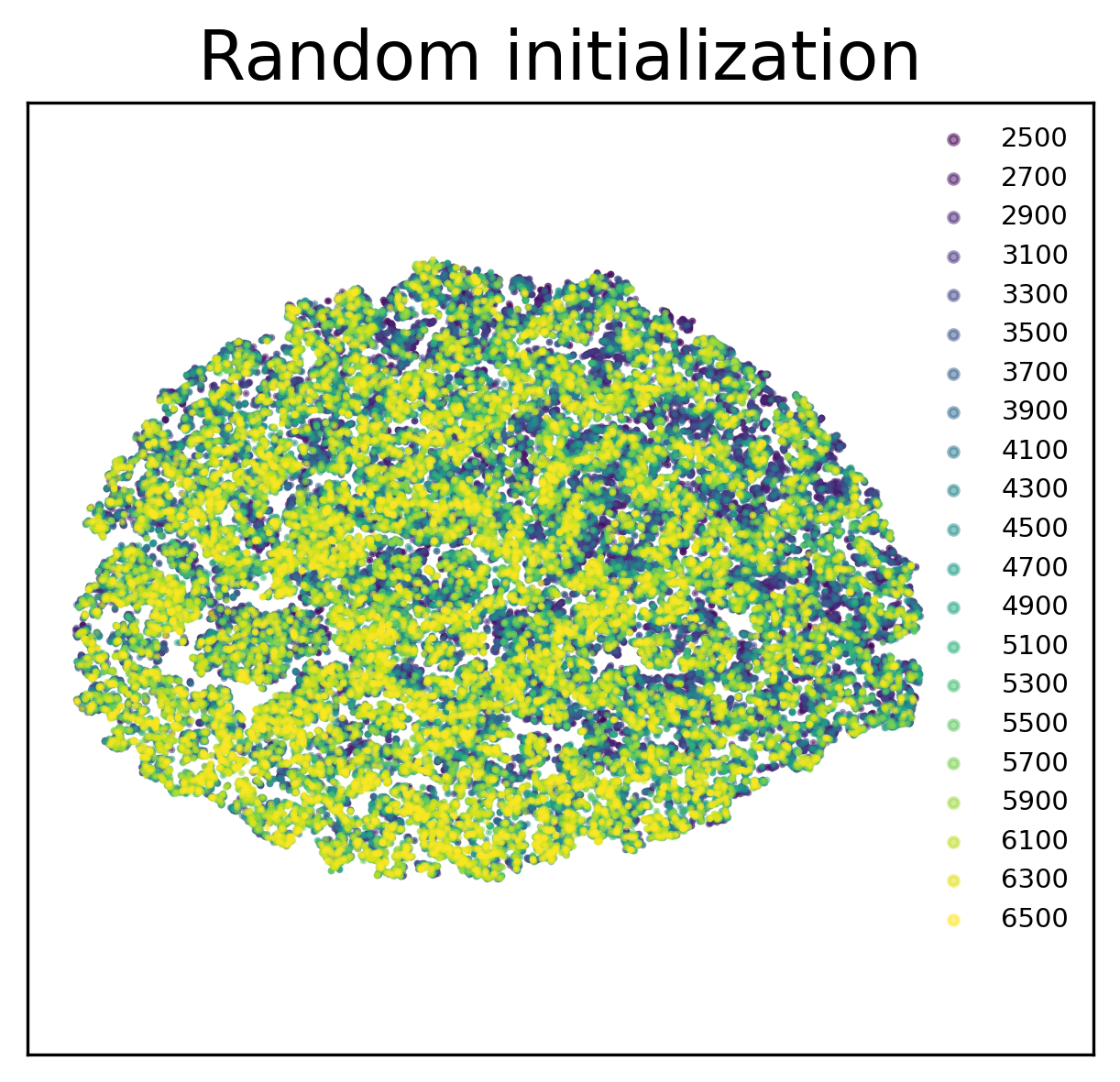}
    \includegraphics[width=0.2\textwidth]{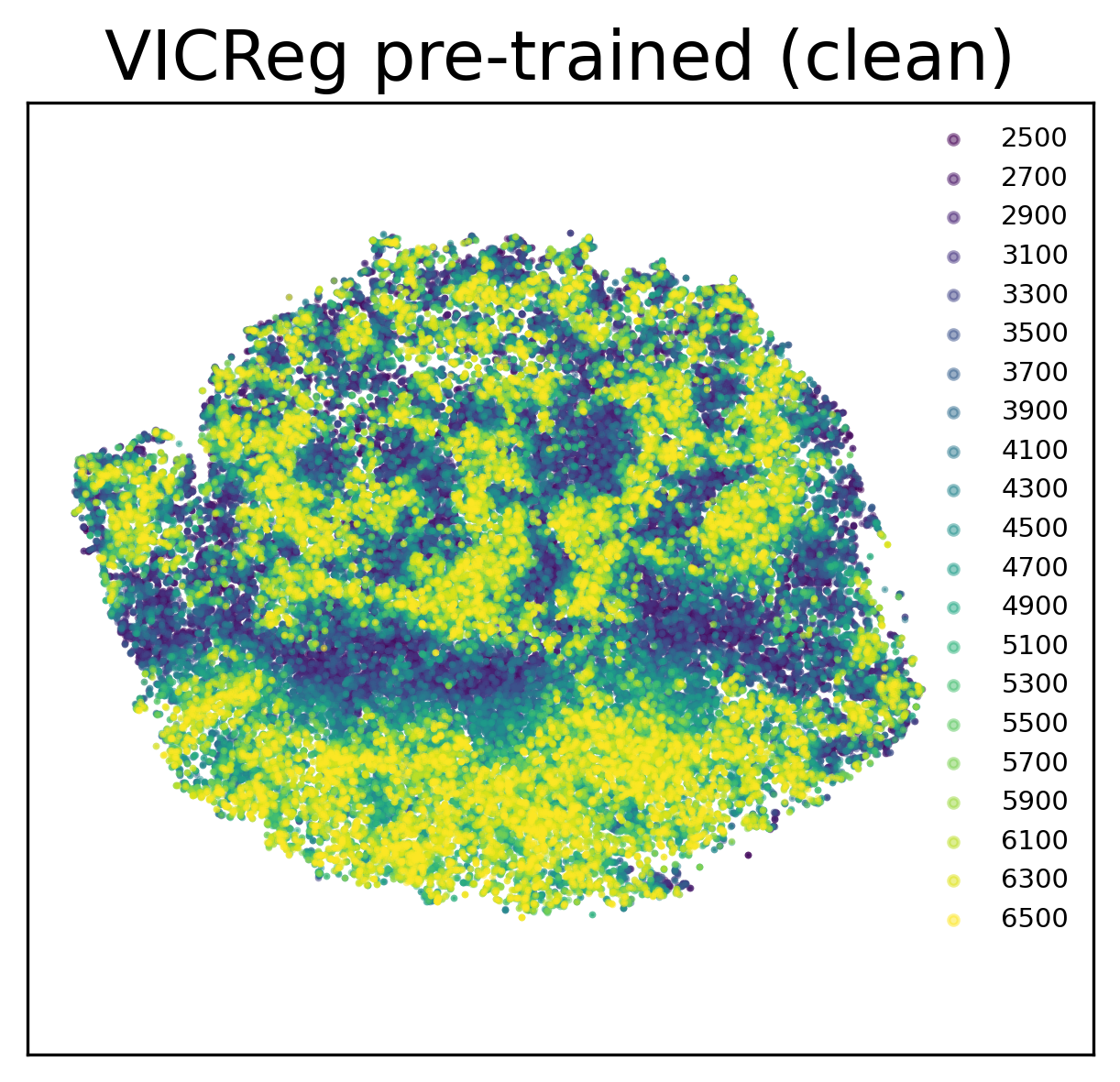}
    \includegraphics[width=0.2\textwidth]{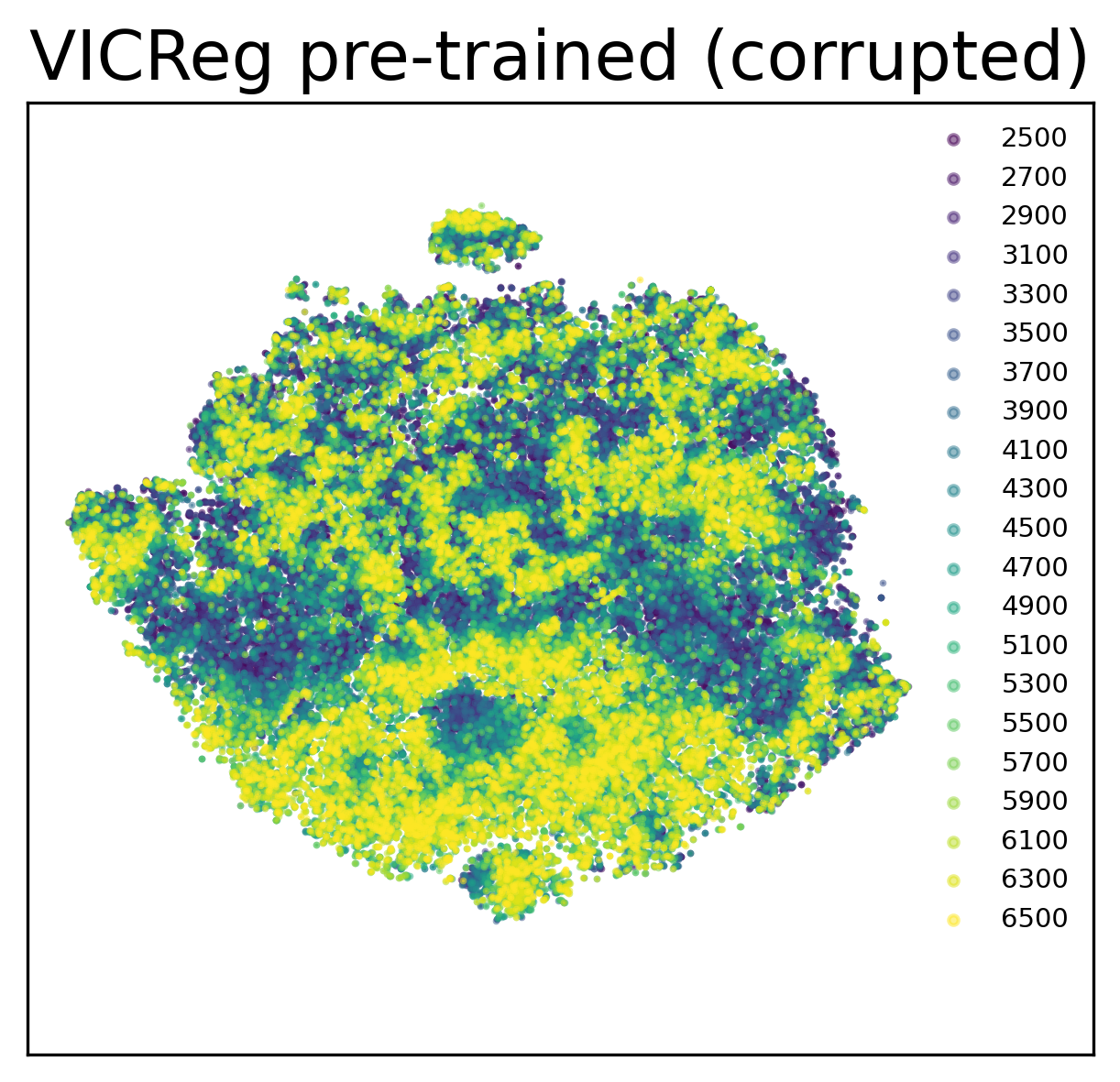}
    \caption{t-SNE visualization of all mass points in the encoder embedding (no mass labels are used for training): random initialization (left), pre-trained encoder on clean data (middle), and nominally corrupted data (right).}
    \label{fig:tsne}
\end{figure}

\textbf{Pre-trained embedding preserved under data corruptions.}
Each of the eleven objects is embedded into a 64-dimensional vector by a dense layer with ReLU activation.
The encoder uses a transformer architecture~\citep{vaswani2017attention}, implemented in Keras~\citep{chollet2015keras}, with three layers, each with four attention heads and feedforward dimension 128 followed by a dropout~\citep{JMLR:v15:srivastava14a} at 0.1.
A 3-layer MLP with hidden dimension 1024 is used as the projection head for the VICReg loss.
The VICReg loss follows the standard weights $\lambda_{\text{inv}}=25$, $\lambda_{\text{var}}=25$, and $\lambda_{\text{cov}}=1$~\citep{bardes2022vicreg}.
The pre-training runs for 300 epochs with batch size 1024, using Adam~\citep{kingma2014adam} with learning rate $10^{-3}$ following a cosine decay~\citep{loshchilov2017sgdr}.
Augmentations are drawn independently per event following Tab.~\ref{tab:aug}.
Fig.~\ref{fig:tsne} shows the 2D projections of the embedding using t-SNE~\citep{JMLR:v9:vandermaaten08a}, comparing before and after the pre-training.
It can be seen that different mass points are well separated and clustered after the pre-training, where no mass labels are used, and the clustering structure is reasonably preserved after combined corruptions are applied to the data, showing robustness of the learned embedding against various systematic variations.

\textbf{Fine-tuning performance.}
We attach a small 3-layer MLP (128, 64, 1) to the pre-trained encoder and train on data with mass labels for relative MSE, for 80 epochs at learning rate $2\times 10^{-4}$, with the encoder weights frozen for the first 15 epochs.
The labeled data has the same set of augmentations applied as in pre-training.
The supervised baseline shares the same architecture and augmentations and is trained from scratch.
Tab.~\ref{tab:relmse} shows that the VICReg model has a better performance with a lower relative MSE on the test set than the supervised model across all corruption scenarios, where the scenarios are defined in Tab.~\ref{tab:aug_eva} in the supplementary material.

\begin{table}[!t]
  \caption{Relative MSE on the test set with different corruption scenarios (mean $\pm$ std over ten independent end-to-end trainings).}
  \label{tab:relmse}
  \centering
  \small
  \scalebox{0.85}{
  \begin{tabular}{lcc}
    \toprule
    Corruptions & Supervised & VICReg \\ \midrule
    Clean (truth) & 0.00317 $\pm$ 0.00047 & \textbf{0.00217 $\pm$ 0.00014} \\
    Particle drop 0.2 & 0.01612 $\pm$ 0.00162 & \textbf{0.01321 $\pm$ 0.00038} \\
    $p_{\text{T}}$ smear 0.2 & 0.00729 $\pm$ 0.00045 & \textbf{0.00626 $\pm$ 0.00006} \\
    $\eta$ smear 0.2 & 0.00496 $\pm$ 0.00066 & \textbf{0.00401 $\pm$ 0.00029} \\
    PID swap 0.2 & 0.00563 $\pm$ 0.00145 & \textbf{0.00397 $\pm$ 0.00024} \\
    Combined (nominal) & 0.01086 $\pm$ 0.00096 & \textbf{0.00914 $\pm$ 0.00020} \\
    \bottomrule
  \end{tabular}
  }
\end{table}

\begin{figure}[!t]
    \centering
    \includegraphics[width=0.161\textwidth]{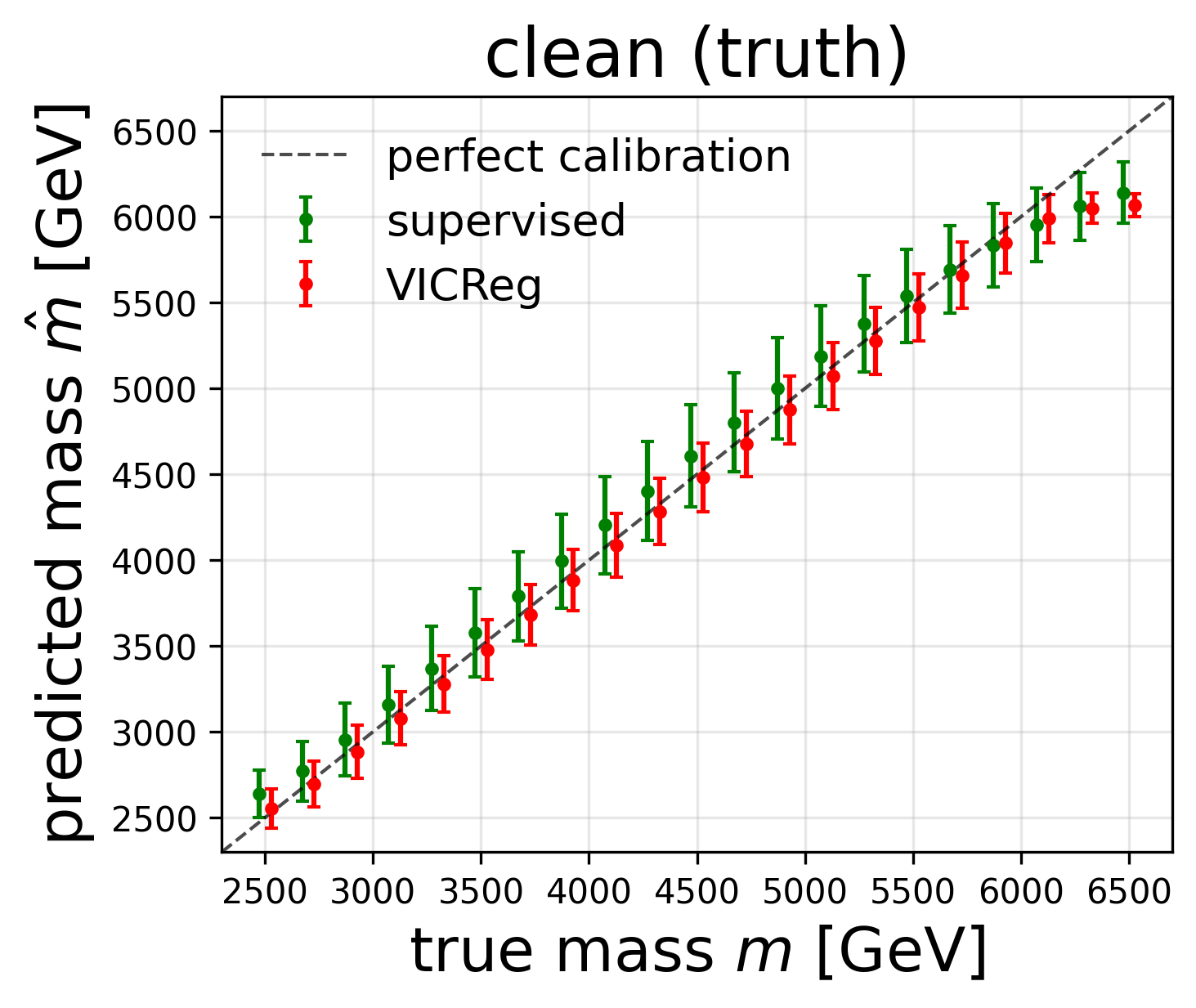}
    \includegraphics[width=0.161\textwidth]{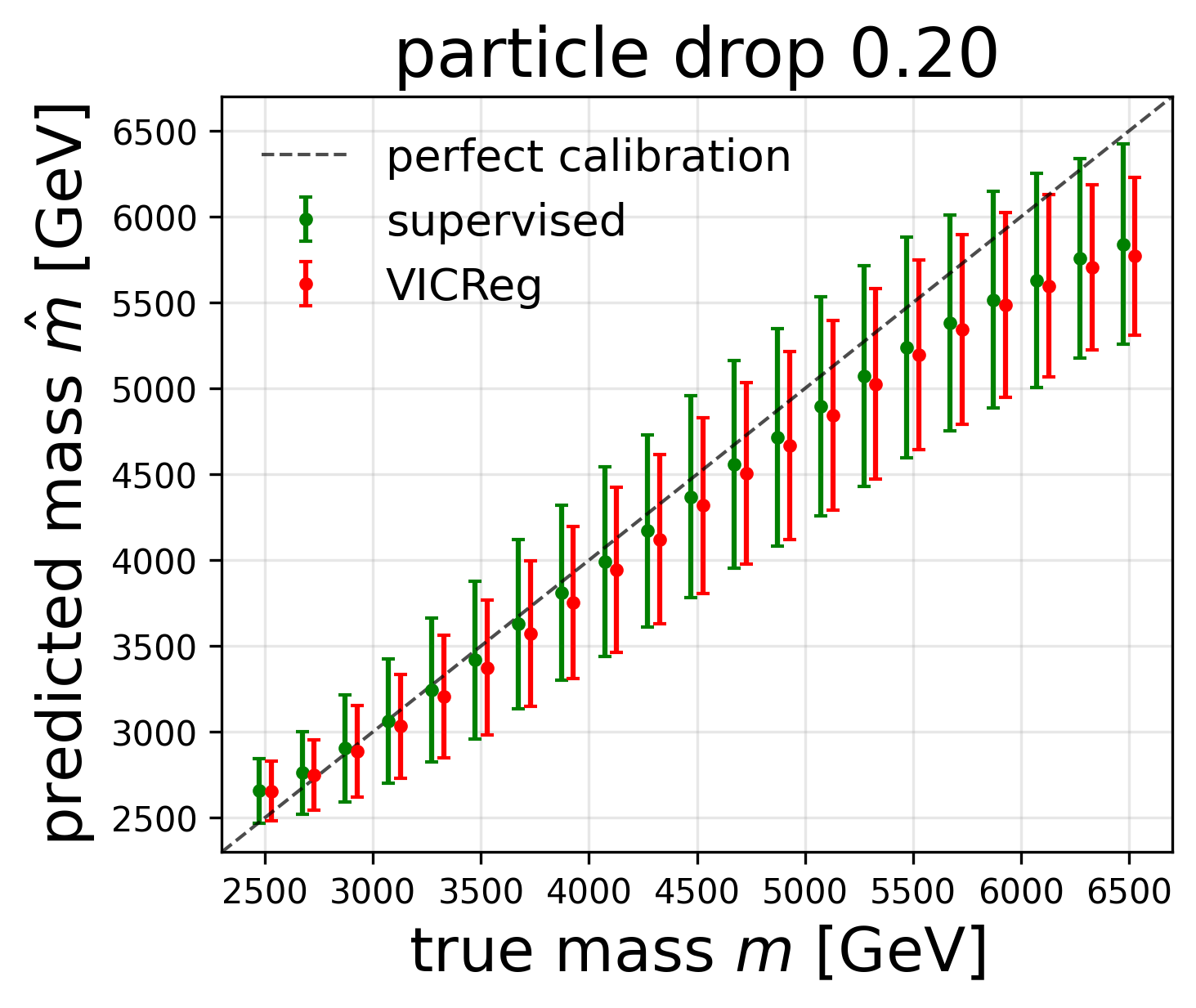}
    \includegraphics[width=0.161\textwidth]{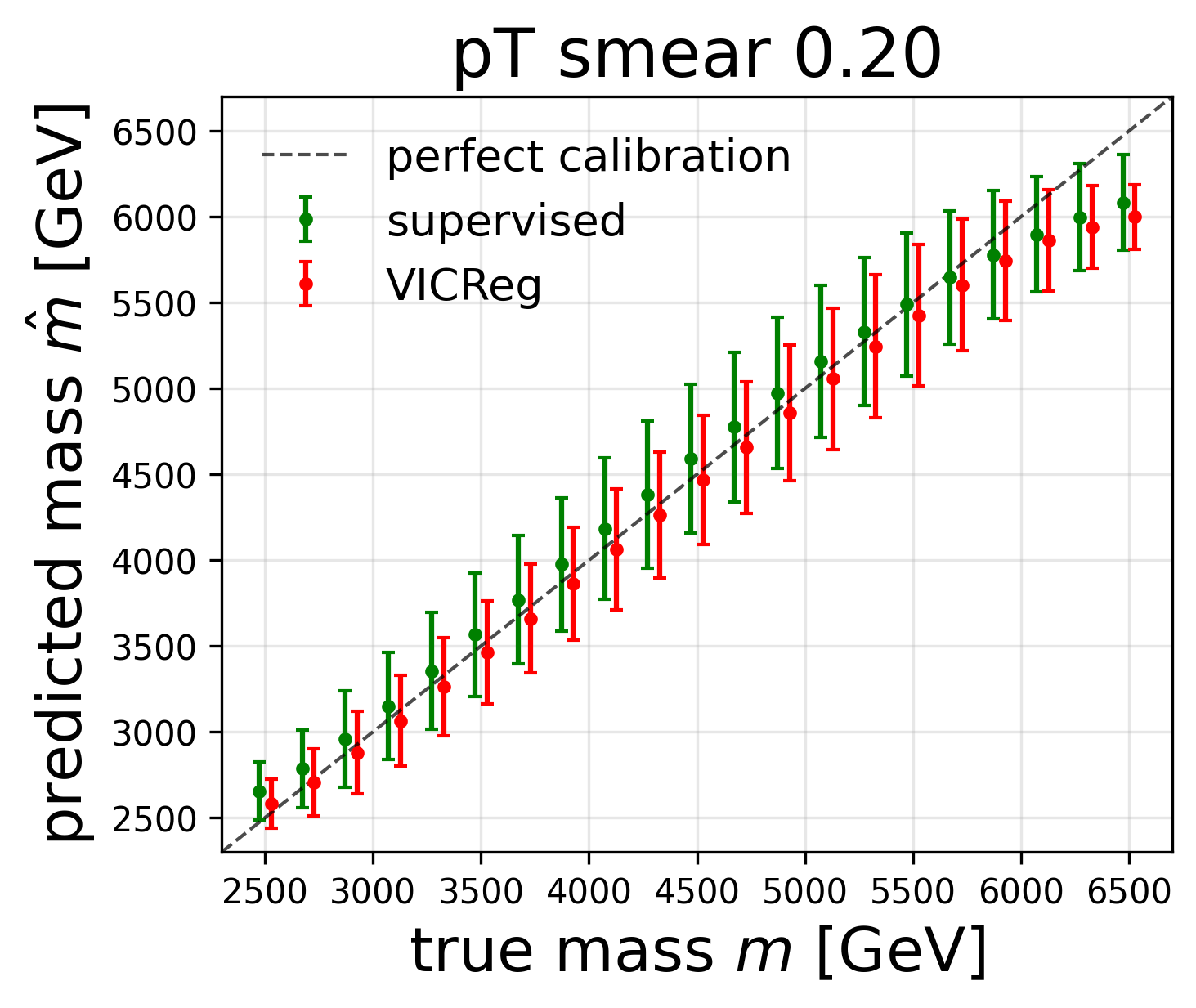}
    \includegraphics[width=0.161\textwidth]{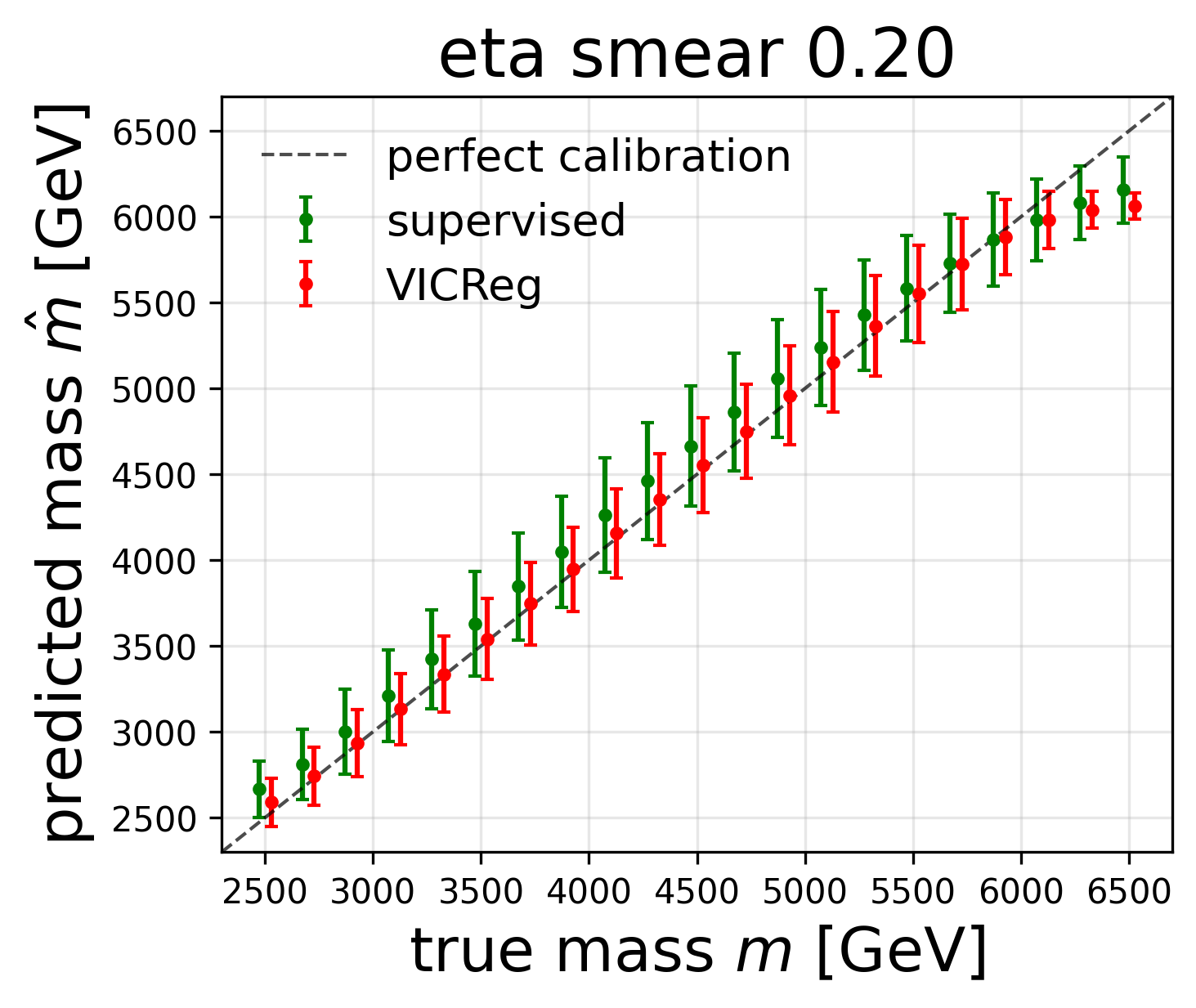}
    \includegraphics[width=0.161\textwidth]{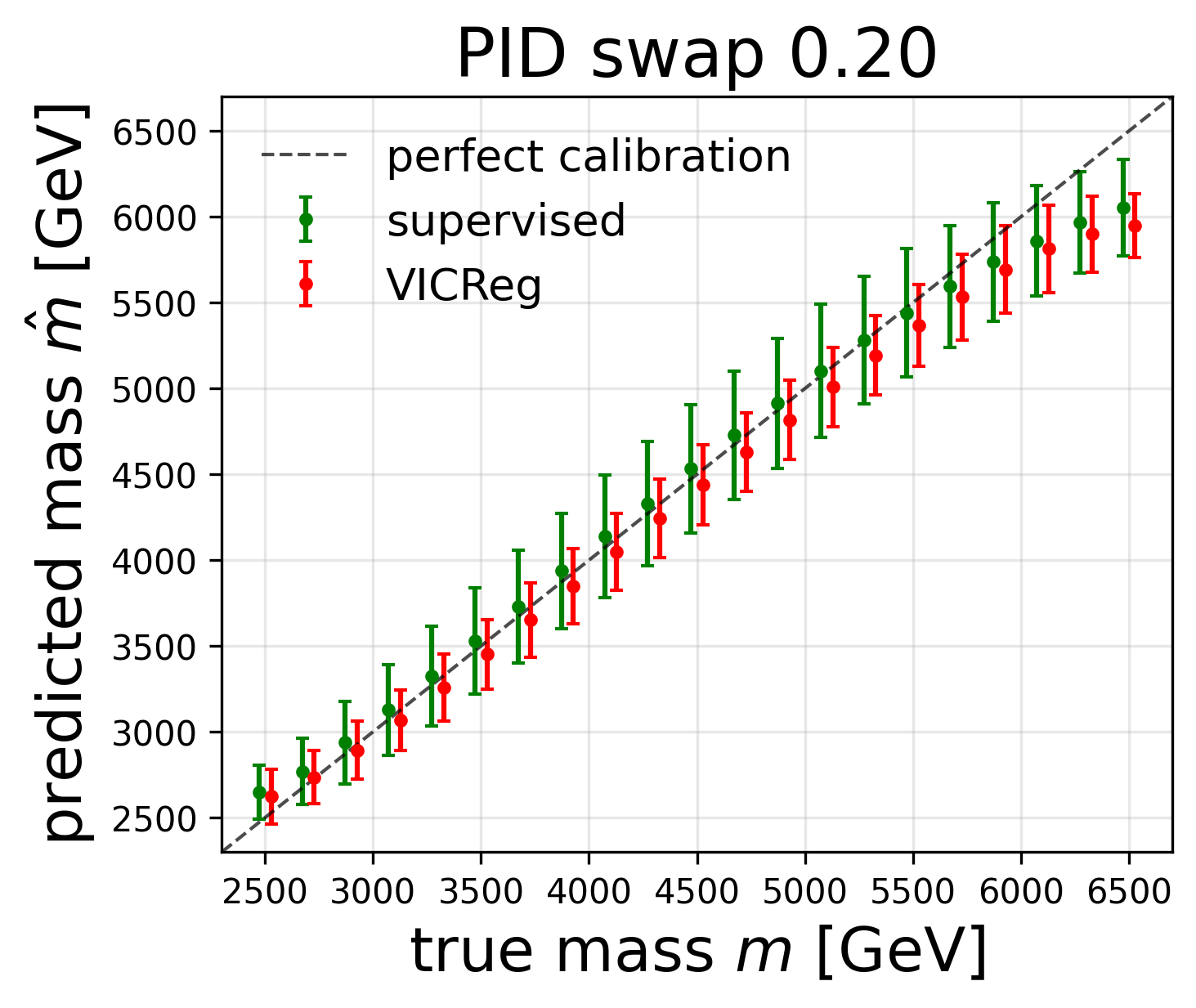}
    \includegraphics[width=0.161\textwidth]{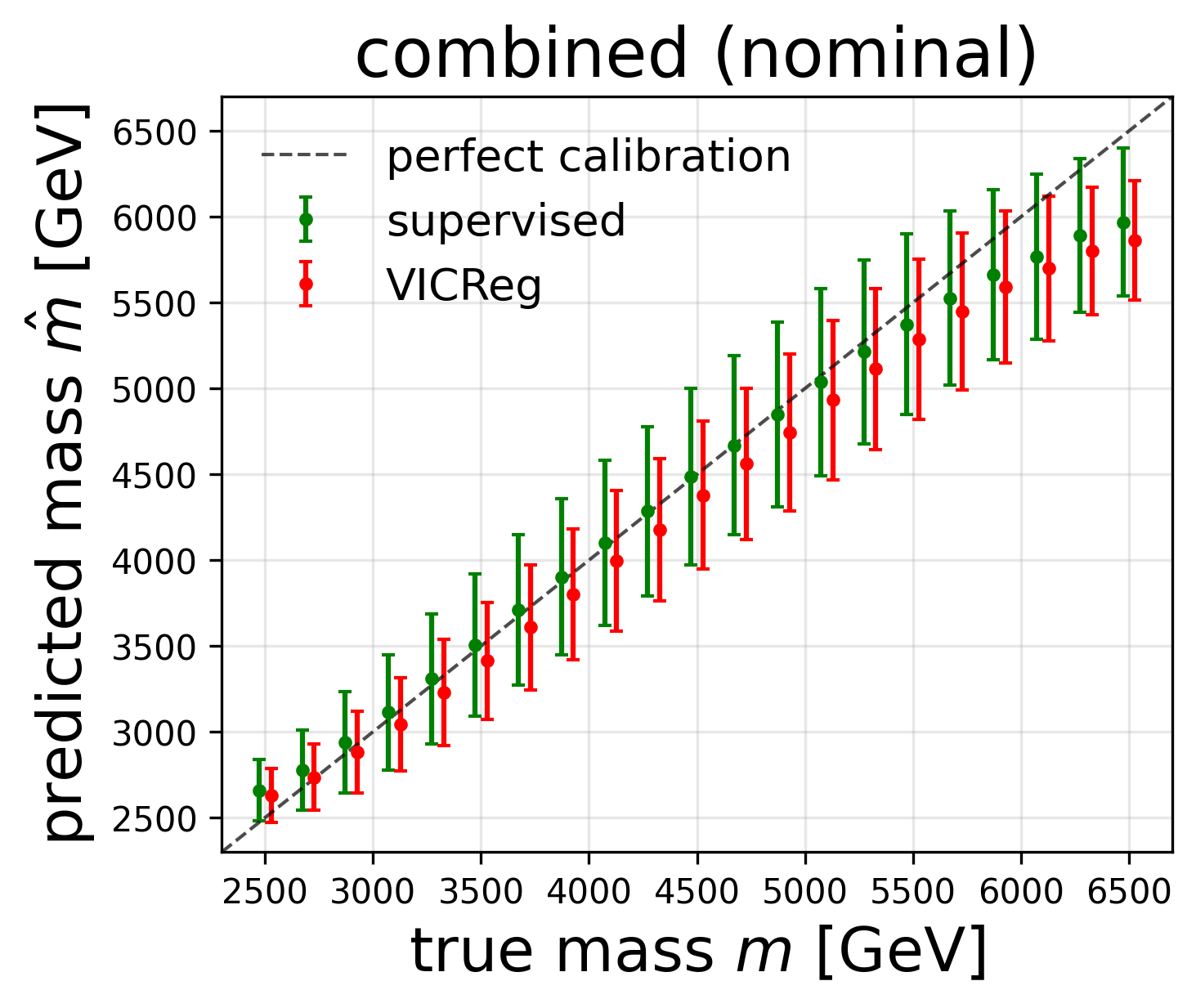}
    \caption{Predicted mass vs. true mass. Error bars are $\pm1\sigma$ of the prediction.}
    \label{fig:permass_calibration}
\end{figure}

\begin{figure}[!t]
    \centering
    \includegraphics[width=0.161\textwidth]{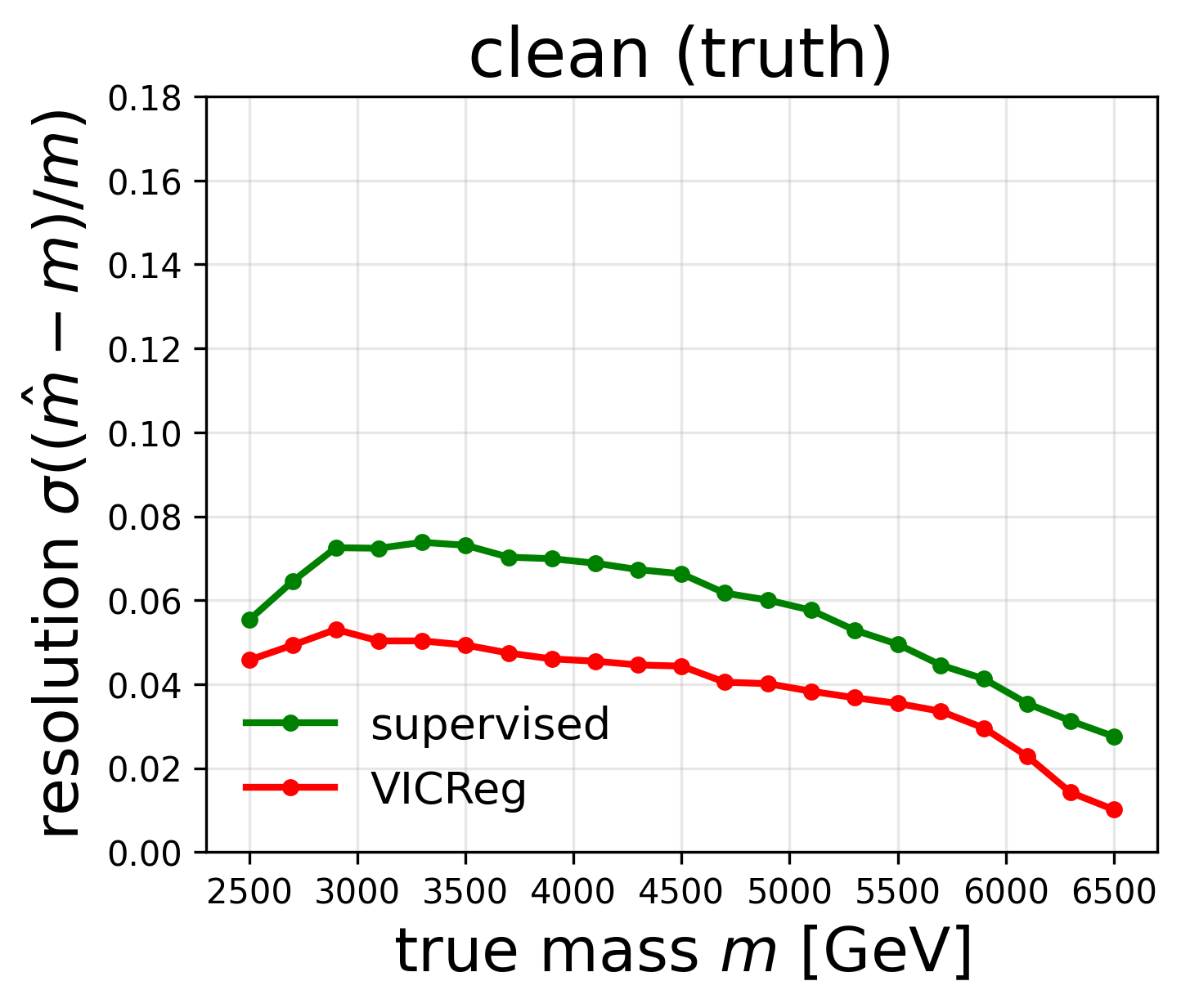}
    \includegraphics[width=0.161\textwidth]{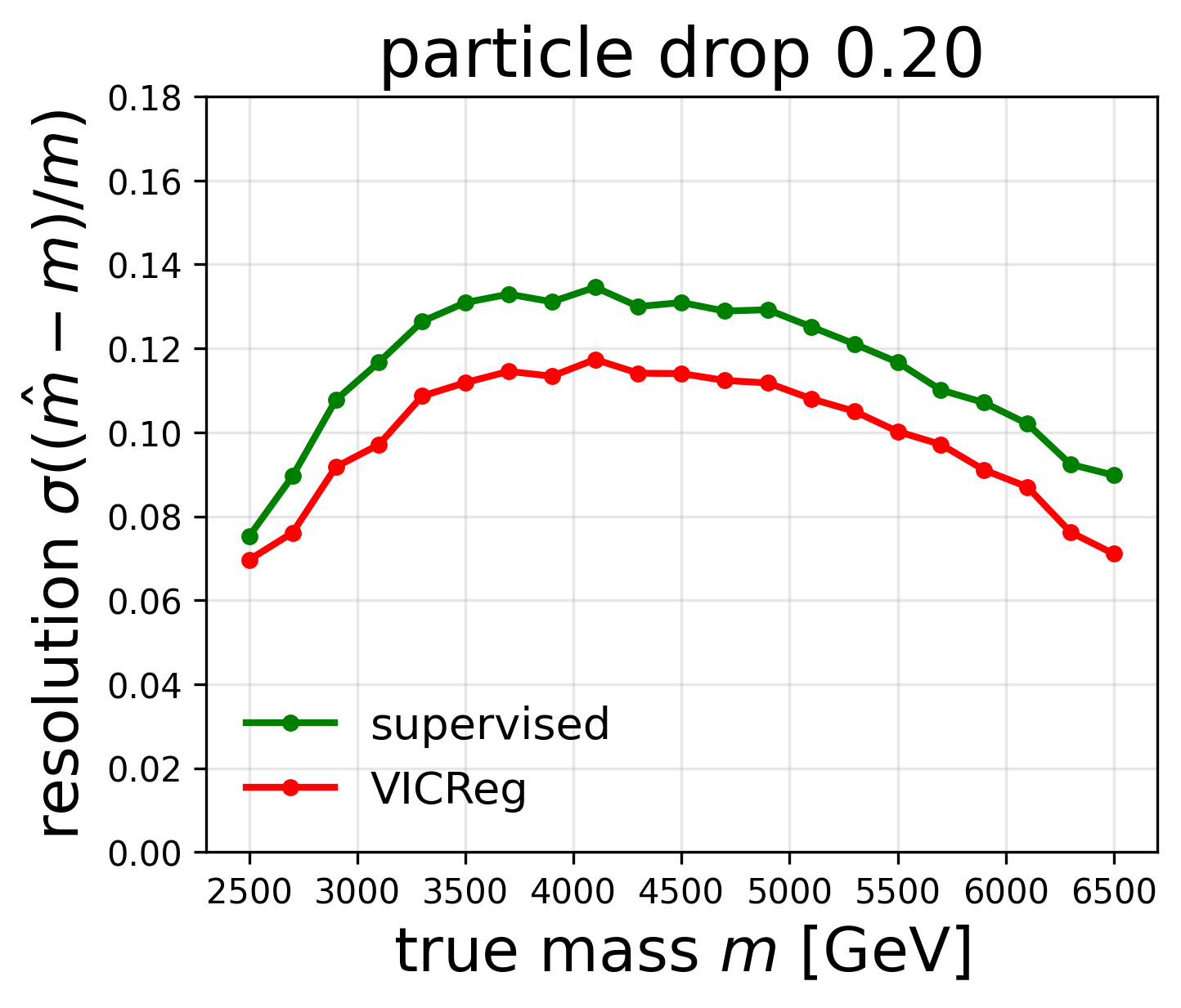}
    \includegraphics[width=0.161\textwidth]{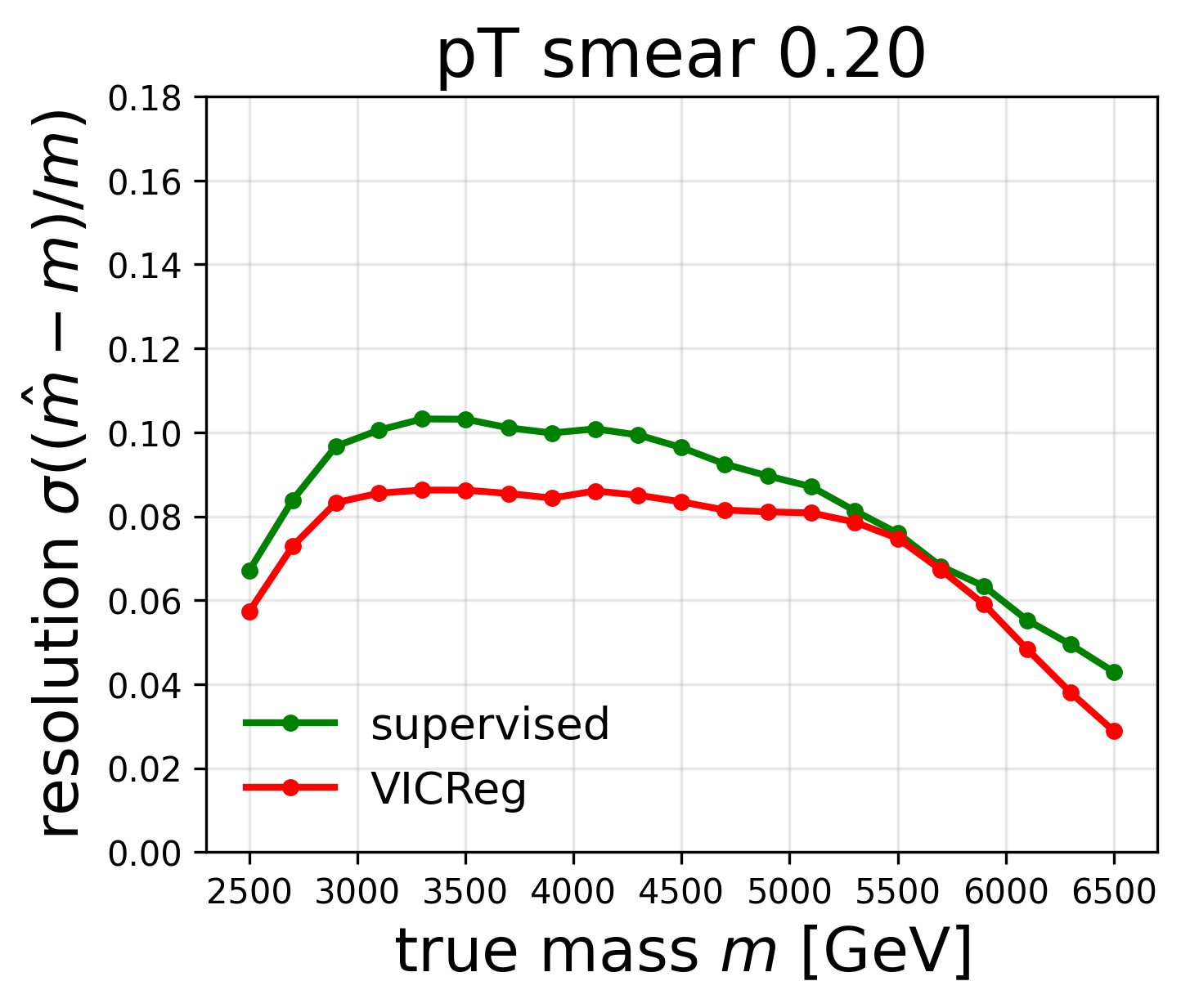}
    \includegraphics[width=0.161\textwidth]{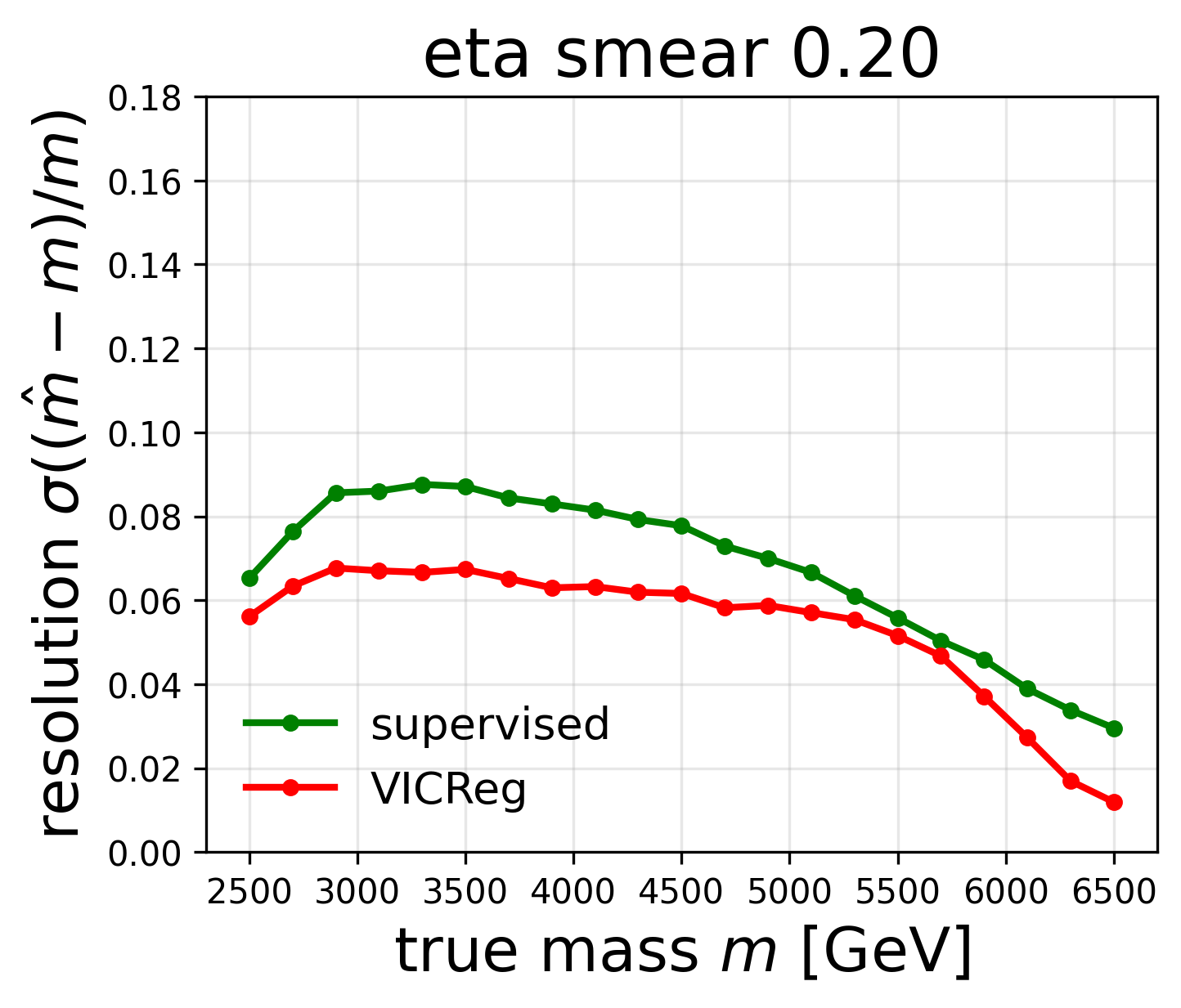}
    \includegraphics[width=0.161\textwidth]{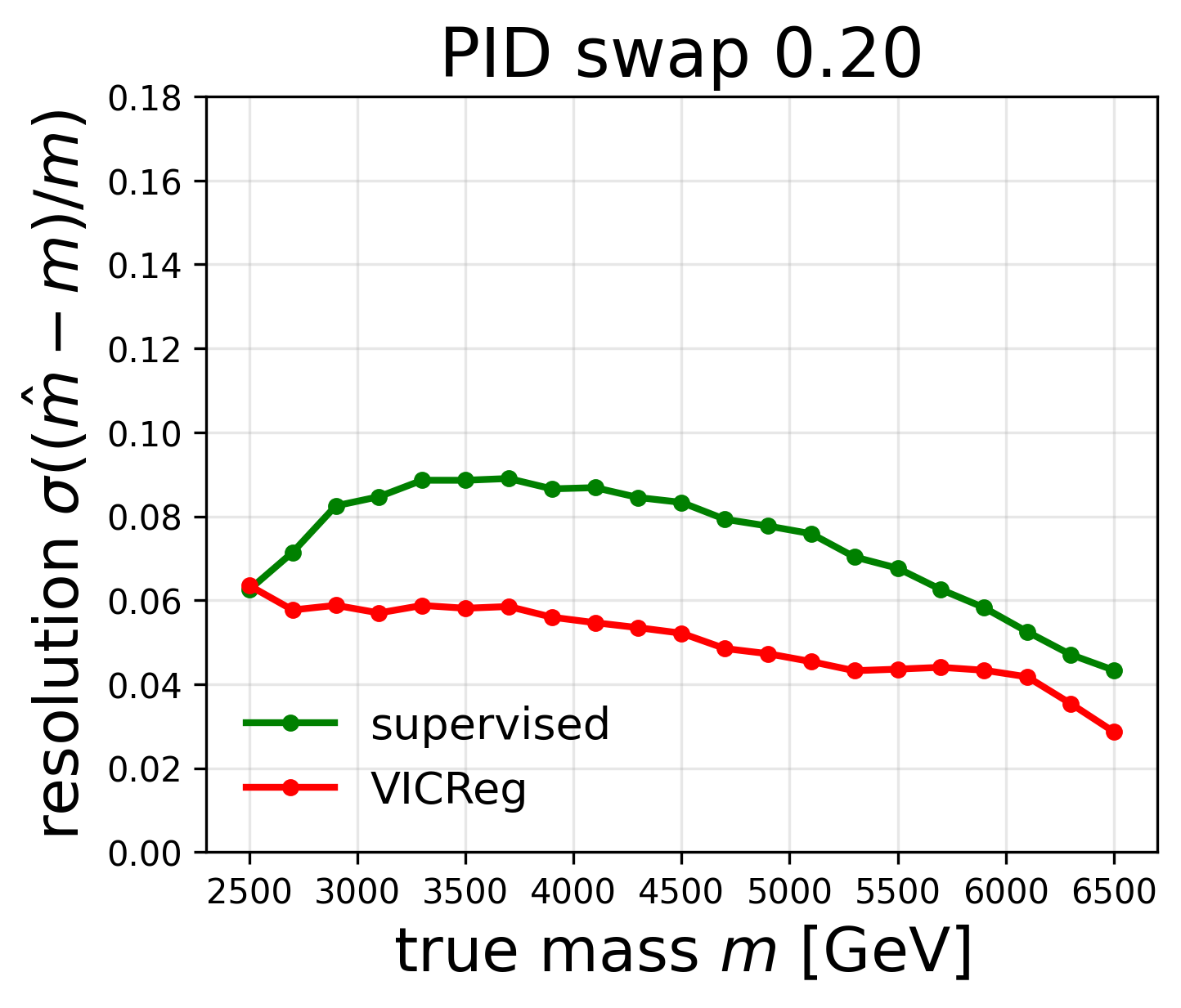}
    \includegraphics[width=0.161\textwidth]{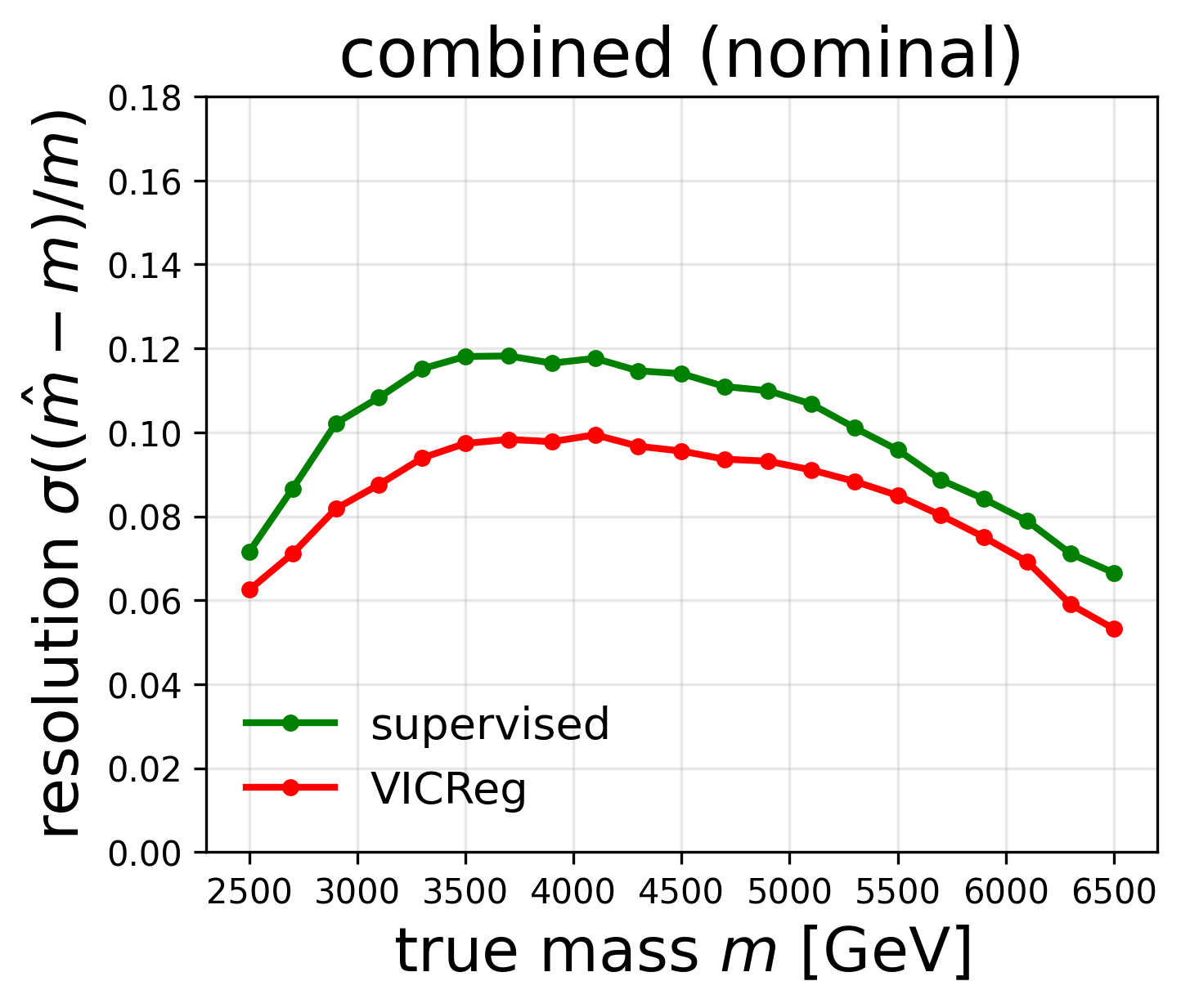}
    \caption{Standard deviation of the mass residual $(\hat{m}-m)/m$ vs. true mass.}
    \label{fig:permass_resolution}
\end{figure}

\begin{figure}[!t]
    \centering
    \includegraphics[width=1\textwidth]{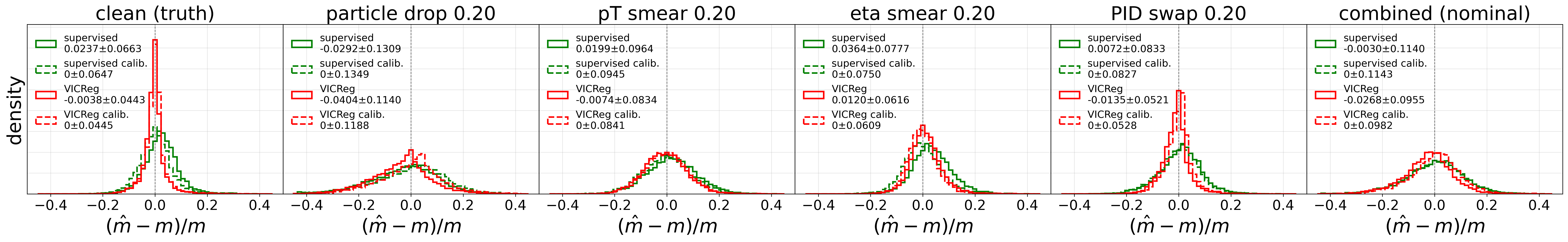}
    \caption{Predicted residual $(\hat{m}-m)/m$ distributions for $m=4500$ GeV. Solid: uncalibrated. Dashed: calibrated to correct the mean residual to zero. The mean and standard deviation are quoted.}
    \label{fig:hist_residual}
\end{figure}

\textbf{Improved resonance widths across corruption types.}
Fig.~\ref{fig:permass_calibration} shows the predicted mass vs. true mass across corruption types.
In the combined corruption scenario, the supervised model predicts the mean slightly closer to the truth than the VICReg model.
We note that this type of response bias can be calibrated away by applying a correction factor to correct the trend, as done in standard regression in HEP analysis~\citep{ATLAS:2023zca}.
However, the predicted mass width cannot be calibrated away, and the VICReg model predicts improved resolution across all corruption scenarios and essentially all mass points, as seen in Fig.~\ref{fig:permass_resolution}, which plots the standard deviation of the residual $(\hat{m}-m)/m$ for each mass point, where VICReg is consistently better than the supervised model with and without corruptions.
These improvements can be attributed to the invariance learned in the pre-training, where events of the same mass are already clustered in the embedding.
The fine-tuned model then inherits this stability, leading to a narrower predicted width at the cost of a small response bias in the mean.
Fig.~\ref{fig:hist_residual} shows the residual distribution for $m=4500$ GeV, before and after a response calibration is applied to each model, showing that the VICReg model narrows the width by between 11\% and 36\% after the calibration depending on the corruption scenario.

\begin{figure}[!t]
    \centering
    \includegraphics[width=0.195\textwidth]{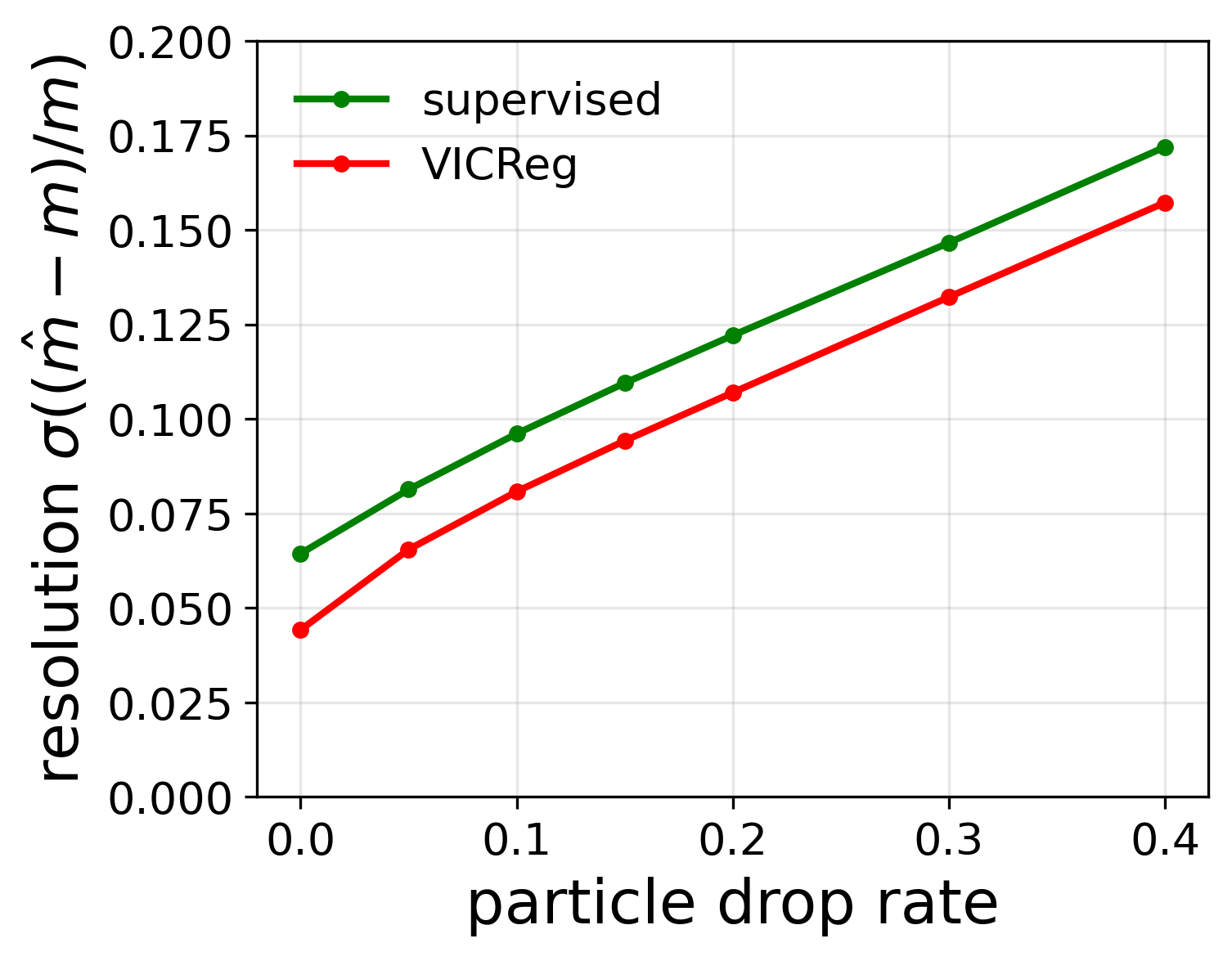}
    \includegraphics[width=0.195\textwidth]{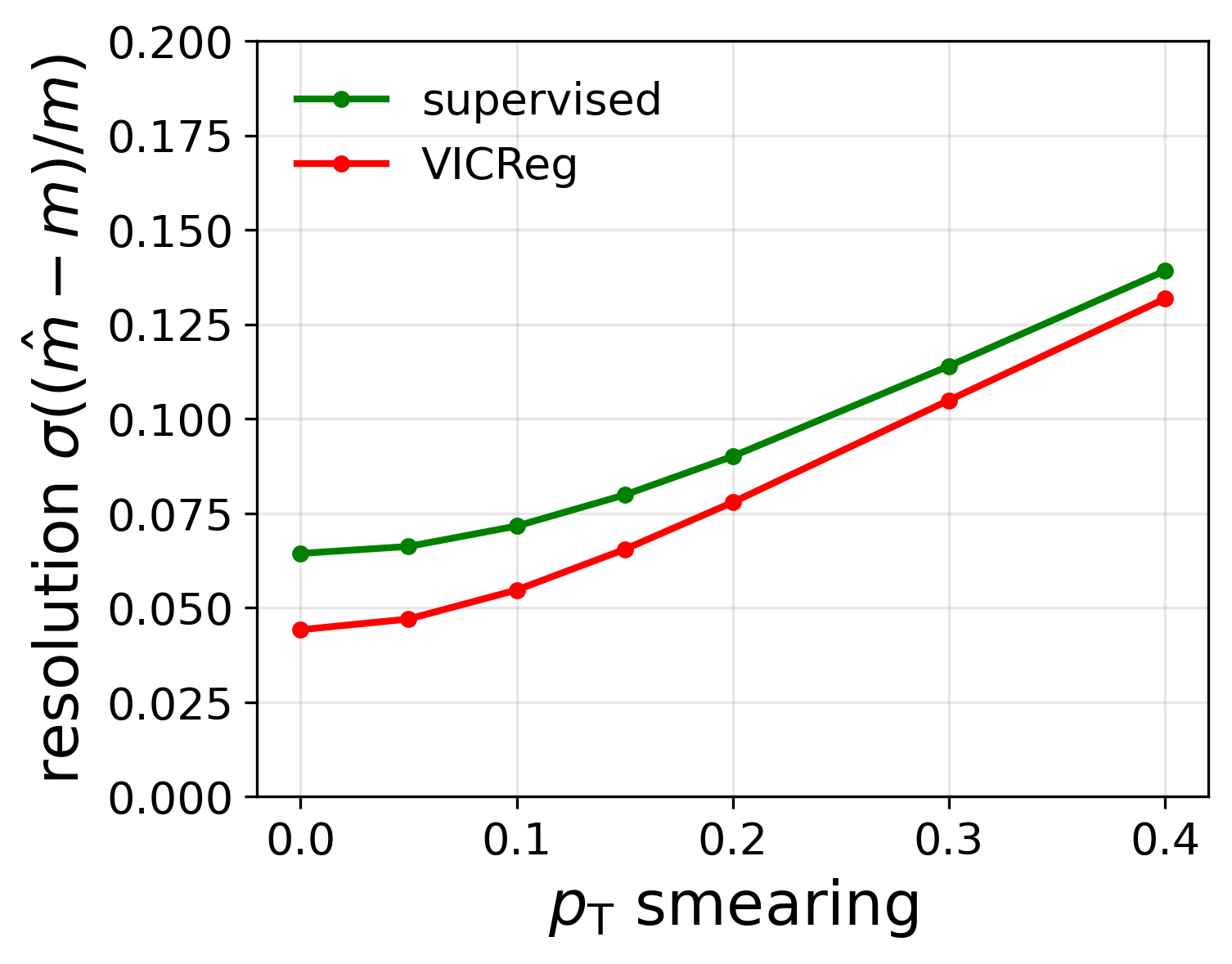}
    \includegraphics[width=0.195\textwidth]{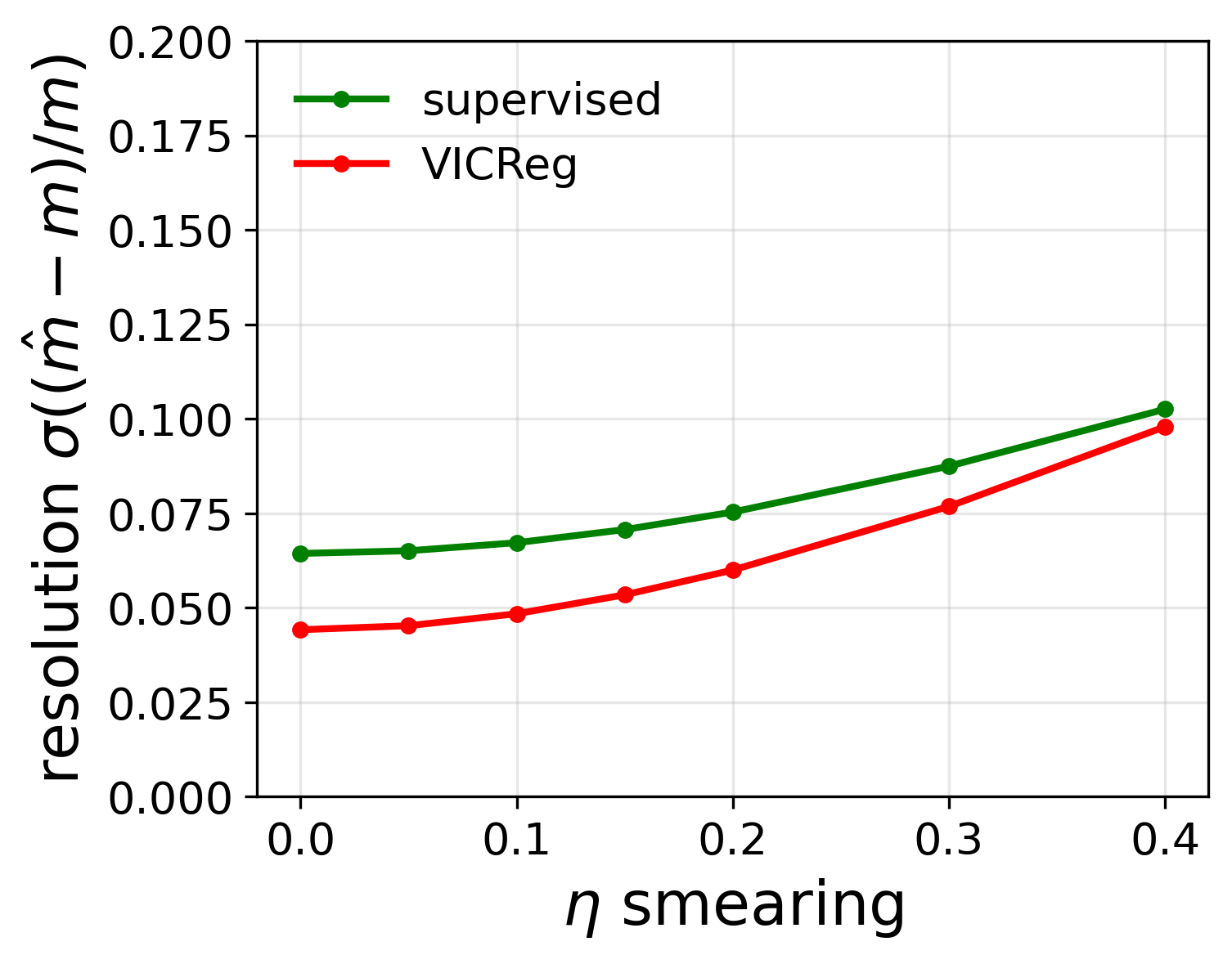}
    \includegraphics[width=0.195\textwidth]{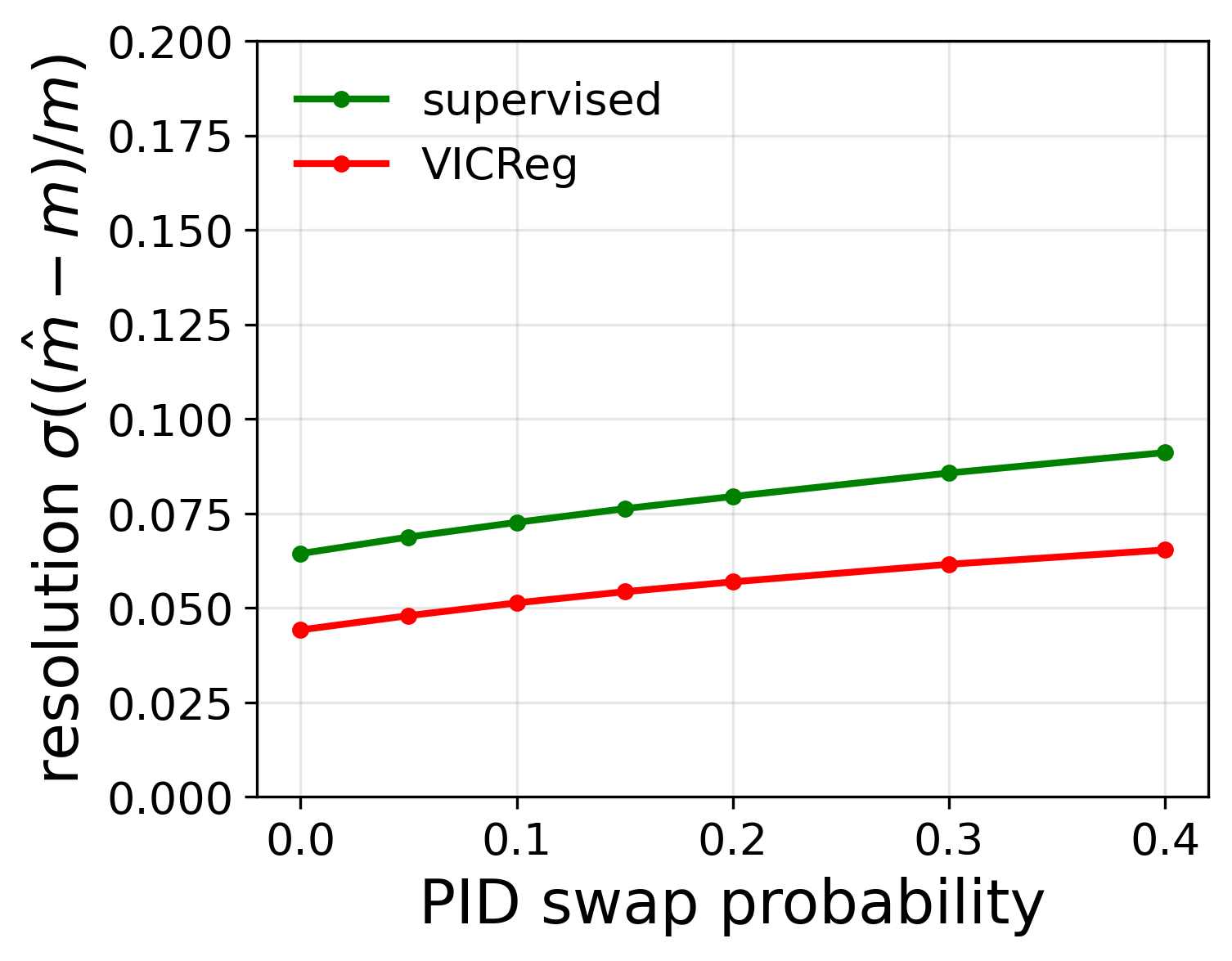}
    \includegraphics[width=0.195\textwidth]{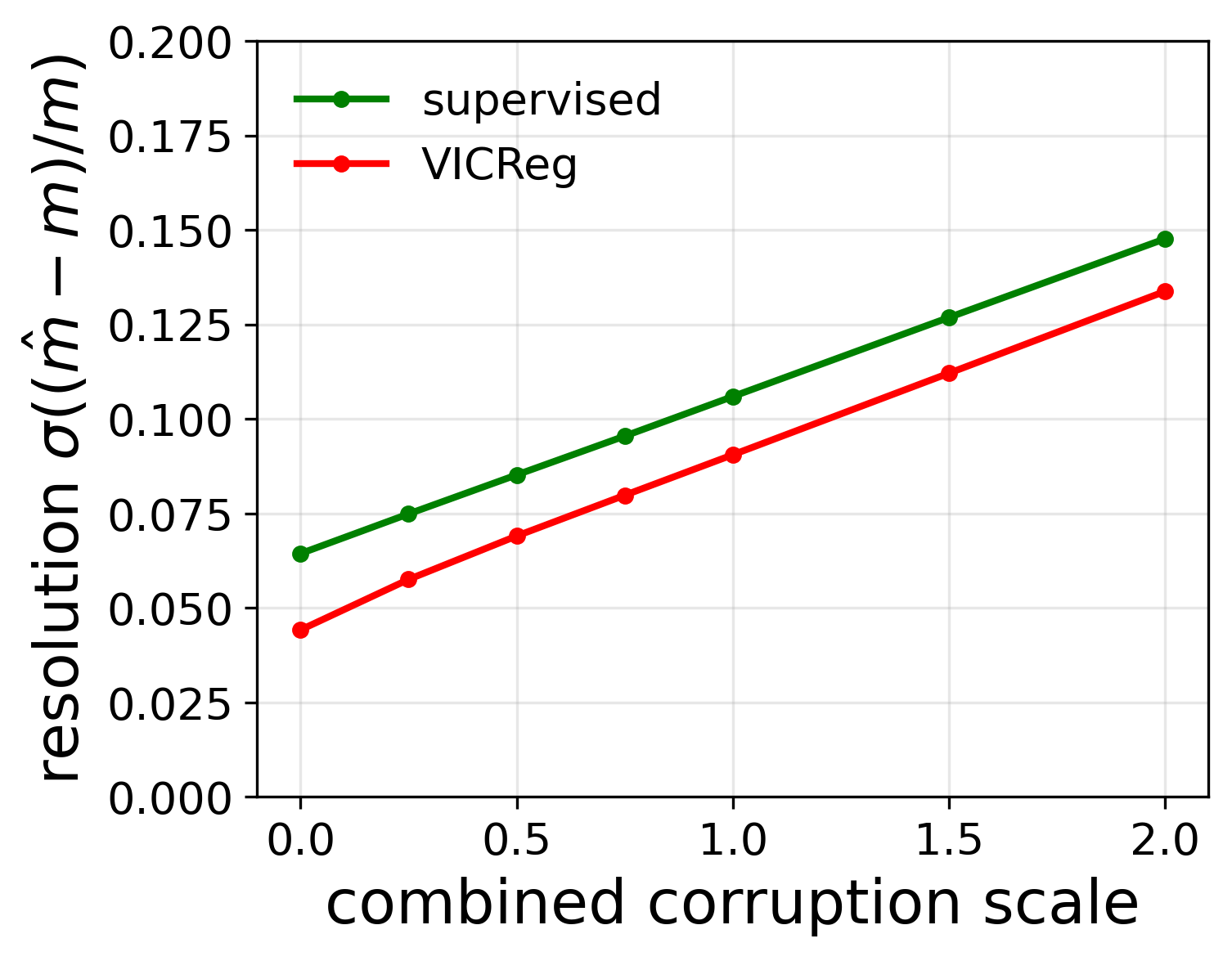}
    \caption{Standard deviation of $(\hat{m}-m)/m$ vs. corruption level.}
    \label{fig:corruption_resolution}
\end{figure}

\textbf{Robustness across corruption levels.}
We also vary the corruption level across different scenarios, where the VICReg model predicts narrower widths compared to the supervised model as shown in Fig.~\ref{fig:corruption_resolution}, showing robustness against increasing corruptions due to the VICReg pre-training.

\section{Conclusion}
We have explored the use of a SSL approach to improve the resolution of reconstructed resonance masses in a way that is robust against various systematic variations, compared to a supervised baseline.
Using a dataset of a TeV-scale heavy resonance with a SUSY-like cascade decay process with eleven final-state objects, we first pre-trained an encoder with the VICReg method to learn an embedding invariant to various realistic systematic variations such as detector mismodeling and reconstruction inefficiencies.
The resulting embedding space shows separation between different mass points even though no mass labels are provided during pre-training.
We then fine-tuned the pre-trained encoder to perform a dedicated mass regression and demonstrated that the VICReg model is able to predict the mass with improved resolution compared to a supervised model across essentially all mass points and all corruption scenarios.
Our results demonstrate the potential of this approach, which yields systematically narrower reconstructed mass widths that are critical for search sensitivity.
After calibrating the response for our model and the supervised baseline, the width of the reconstructed mass is improved with SSL by roughly 14\% when applying all corruption types.
Future directions include scaling to larger datasets with full simulation chains, exploring more realistic corruption schemes, and using a wider range of regression targets for a systematic probe of the approach.

\begin{ack}
HFT and DSR are supported by the U.S. Department of Energy (DOE), Office of Science, Office of High Energy Physics subprogram on Computational High Energy Physics under Award No. DE-SC0026801.
This work used resources available through the National Research Platform (NRP) at the University of California, San Diego~\citep{10.1145/3708035.3736060}.
NRP has been developed, and is supported in part, by funding from National Science Foundation, from awards 1730158, 1540112, 1541349, 1826967, 2112167, 2100237, and 2120019, as well as additional funding from community partners.
Generative AI (Anthropic's Claude) was used to assist with coding and language editing.
\end{ack}


\bibliographystyle{plainnat}
\bibliography{references}


\clearpage


\appendix

\section{Supplementary Material}

\textbf{Corruption scenarios for evaluation.} Tab.~\ref{tab:aug_eva} lists the corruption scenarios for test set evaluation, with the same labels shared across Tab.~\ref{tab:relmse} and other evaluation figures.
These evaluation corruptions are harsher than the augmentations used for pre-training (Tab.~\ref{tab:aug}), as they are applied at equal or higher strengths without event-level probabilities in order to probe the generalization and robustness of the embedding beyond augmentations seen during pre-training.

\begin{table}[!t]
  \caption{Corruption scenarios for evaluation.}
  \label{tab:aug_eva}
  \centering
  \small
  \scalebox{0.85}{
  \begin{tabular}{ll}
    \toprule
    Scenario & Augmentation applied to test events \\ \midrule
    Clean (truth) & None \\
    Particle drop 0.2 & Each object removed with prob. 0.2 \\
    $p_{\text{T}}$ smear 0.2 & All objects have a $p_{\text{T}}$ smearing at $\sigma=20\%$ \\
    $\eta$ smear 0.2 & All visible objects have an $\eta$ smearing at $\sigma=0.2$ \\
    PID swap 0.2 & Each object's PID replaced by that of a random object with prob. 0.2 \\
    Combined (nominal) & All of the above at 0.1 / 0.1 / 0.1 / 0.05 \\
    \bottomrule
  \end{tabular}
  }
\end{table}

\begin{figure}[!t]
    \centering
    \includegraphics[width=0.4\textwidth]{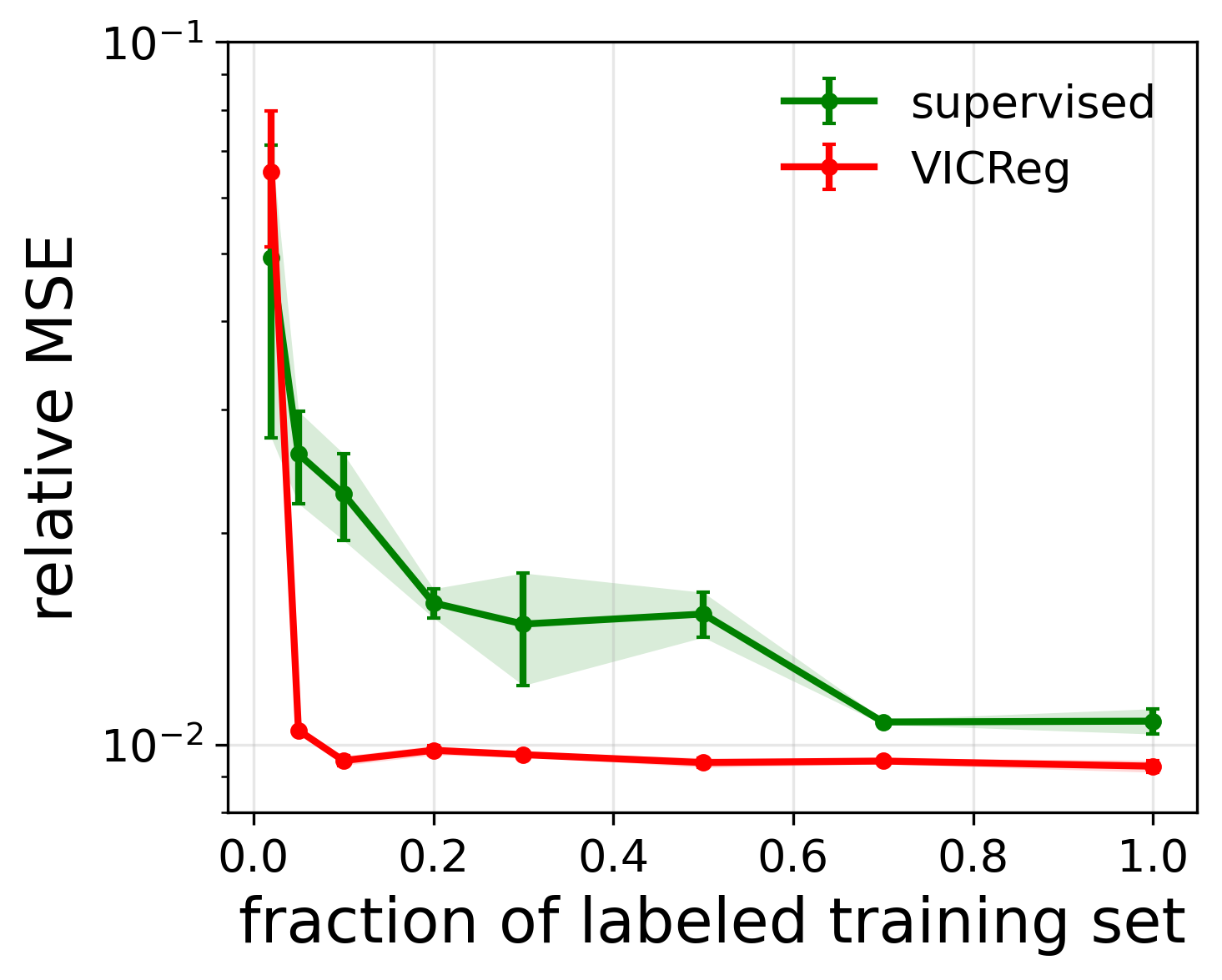}
    \caption{Relative MSE on the test set for models trained at different train set sizes, evaluated at the combined (nominal) corruption.}
    \label{fig:sample_eff}
\end{figure}

\textbf{Labeled sample efficiency.} Fig.~\ref{fig:sample_eff} shows that the VICReg model has a higher sample efficiency, converging more stably at the lowest relative MSE loss when fine-tuned on only around 10\% of labeled data, while the supervised model needs 70\% to converge, demonstrating that the VICReg model requires much less labeled data to train for best performance.
This feature is valuable as simulating labeled data in HEP is computationally expensive.

\end{document}